\documentclass[twocolumn,pra,showpacs]{revtex4}
\usepackage{amsmath}
\usepackage{amssymb}
\usepackage{graphicx}
\usepackage{amsfonts}
\usepackage{txfonts}



\begin{document}

\title{Proposal for geometric generation of a biexciton in a quantum dot
using a chirped pulse}

\author{H. Y. Hui}
\author{R. B. Liu}
\affiliation{Department of Physics, The Chinese University of Hong Kong, Shatin, Hong Kong, China}

\date{\today}

\begin{abstract}
We propose to create a biexciton by a coherent optical process using a frequency-sweeping (chirped)
laser pulse. In contrast to the two-photon Rabi flop scheme, the present method uses the state
transfer through avoided level crossing and is a geometric control. The proposed process is robust
against pulse area uncertainty, detuning, and dephasing. The speed of the adiabatic operation is
constrained by the biexciton binding energy.
\end{abstract}


\maketitle

\section{Introduction}

Semiconductor quantum dots (QDs) have manifold uses in quantum information
and computation. They have been utilized to generate
single-photons~\cite{michler2000qds,benson2000rae,moreau2001sms,yuan2002eds,santori2001tsp,Flissikowski2001}
with good indistinguishability~\cite{santori2002ips}. More recently
it has been proposed and partially
realized~\cite{benson2000rae,michler2000qca,santori2002pcp,ulrich2003tpc,reimer_voltage_2007}
that QDs in two-exciton states, called biexcitons, can be used to
generate pairs of entangled photons, by cascade emission of
photons~\cite{moreau2001qcp}. The entanglement of photon pairs in this scheme
was noted~\cite{stace2003etp} to be imperfect, because of the slight
difference in energy between the two single-exciton levels~\cite{Gammon1996,kulakovskii1999fsb}.
However, considerable improvement has recently been made~\cite{akopian2006epp,stevenson2006sst},
which suggests that the scheme should be of high experimental value
in quantum optics, quantum computation~\cite{knill2001seq} and quantum
cryptography~\cite{ekert1991qcb,jennewein2000qce}, and can also
be used to test foundations of quantum mechanics~\cite{benson2000rae}.
Biexciton is also of interest in itself because it serves as the physical
basis for a 2-bit conditional quantum logic gate~\cite{li2003aoq}.

A number of works have already been done on the optical coherent control
of the single-exciton states in, e.g., InAs/GaAs
QDs~\cite{bonadeo1998coc,htoon2002iro,kamada2001ero,stievater2001roe,zrenner2002cpt}
and CdSe/ZnSe QDs~\cite{Flissikowski2001}. In the recent experiments
of optical coherent control of biexciton states, two approaches were
used. The first one applies two optical beams each in resonance with
the $\left|g\right\rangle \rightarrow\left|X\right\rangle $ (ground
state to single exciton) and $\left|X\right\rangle \rightarrow\left|XX\right\rangle $
(single- to bi-exciton) transitions~\cite{chen2002bqc,li2003aoq}.
However, it was noted~\cite{akimov2006seb} that a better approach
is to apply degenerate pulses with frequency equal to half the biexciton
energy, such that the spontaneously emitted photons have frequencies
different from that of the excitation pulse. This has been followed
by recent works~\cite{akimov2006seb,flissikowski2004tpc,stufler2006tpr}.
Experiments have been done on both InAs/GaAs and CdSe/ZnSe QDs, and
the phenomenon of two-photon Rabi oscillation is the prime indicator
of successful control in these experiments.

In this paper, we propose to use a frequency-sweeping pulse~\cite{goswami2003ops}
for a geometric generation of a biexciton in a quantum dot. The scheme
is based on the adiabatic state transfer from the ground state to
the final biexciton state via avoided energy level crossing, in which
the intermediate exciton is bypassed. The utilization of level anti-crossing
follows the idea of the STIRAP (stimulated Raman adiabatic passage)
for adiabatic state transfer in a $\Lambda$-type 3-level system~\cite{bergmann1998cpt,laine1996apt}.
But here since both the exciton and biexciton transitions couple to
the same optical pulse, independent control of the two transitions
as required in the STIRAP is not feasible. Instead, the frequency
sweeping~\cite{goswami2003ops} is proposed to realize the adiabatic
state evolution. The geometric scheme bears the robustness against
some uncertainty in the system parameters such as energy levels and
dipole magnitude and in laser pulse parameters such as amplitude,
shape, and frequency, which is unavoidable in realistic experiments.
Bypassing the intermediate single-exciton state minimizes the possibility
of generating single-photon emission which, e.g., may contaminate
an entangled photon pair in quantum optics application. Constrained
by the biexciton binding energy, the adiabatic state transfer can
be completed in picosecond timescales for a typical CdSe quantum dot,
and thus the effect of the exciton dephasing can be largely avoided.

This paper is organized as follows: In Sec.~\ref{sec:HAMILTONIAN}
we formulate the problem and give the waveform of the pulse used;
in Sec.~\ref{sec:EVOLUTION} we demonstrate numerically the creation
of a biexciton which is robust against small uncertainty in all parameters
characterizing the system and keeps the occupation of single-exciton
state relatively low; in Sec.~\ref{sec:effects-of-dephasing} we
show that dephasing, modeled in the Lindblad formalism~\cite{lindblad1976gqd},
only slightly reduce the efficiency.

\section{Model and mechanism}
\label{sec:HAMILTONIAN}

The biexciton system can be modeled by a four-level system: the ground
state $\left|g\right\rangle $, the biexciton state $\left|XX\right\rangle $,
and two intermediate single-exciton states with different linear polarizations
$\left|X\right\rangle $ and $\left|Y\right\rangle $~\cite{Gammon1996,kulakovskii1999fsb}.
Because the two pathways of excitation, $\left|g\right\rangle \rightarrow\left|X\right\rangle \rightarrow\left|XX\right\rangle $and
$\left|g\right\rangle \rightarrow\left|Y\right\rangle \rightarrow\left|XX\right\rangle $,
is independent and can be implemented independently by applying different
polarizations of excitation in experiments, we only consider $\left|g\right\rangle \rightarrow\left|X\right\rangle \rightarrow\left|XX\right\rangle $.

\begin{figure}[t]
\begin{centering}
\includegraphics[width=7.5cm]{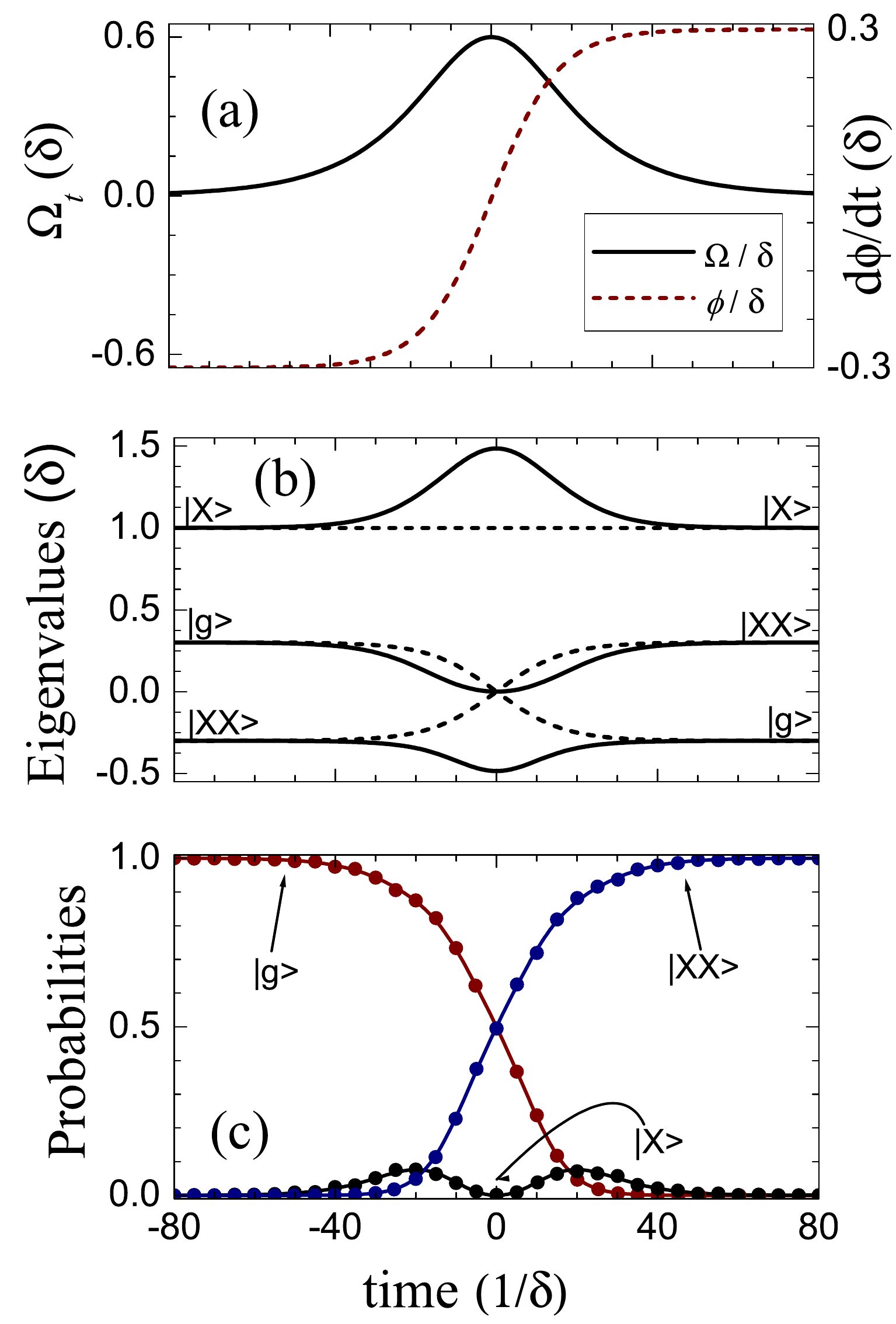}
\caption{(Color online) (a) Laser pulse amplitude and time-dependent frequency
as defined in Eq.~(\ref{eq:pulseshape}): $\Omega_{t}$ and $\dot{\phi}_{t}$,
scaled with respect to $\delta$.\quad{}(b) The three eigenvalues
of Hamiltonian in Eq.~(\ref{eq:Hmod}) under the pulse in Eq.~(\ref{eq:pulseshape})
(solid lines), and when $\Omega_{t}=0$ (dashed lines). The eigenvector
corresponding to each eigenvalue at the beginning and the end of the
pulse is indicated.\quad{}(c) The evolution of the system by adiabatic
approximation (solid lines) and numerical integration of Schr\"{o}dinger
equation (dots), under the initial condition that only $\left|g\right\rangle $
is occupied. The parameters here are $A=0.6\delta$, $\alpha=0.06\delta$,
$\mu=5$, $T=80/\delta$. }
\label{fig:adiabatic}
\par\end{centering}
\end{figure}

The Hamiltonian is written as
\begin{align}
H =& \left(\omega+\delta\right)\left|X\right\rangle \left\langle X\right|+2\omega\left|XX\right\rangle \left\langle XX\right|
\nonumber \\
&+\left[\Omega(t)\left(\left|g\right\rangle \left\langle X\right|+\left|X\right\rangle \left\langle XX\right|\right)+H.c.\right],
\label{eq:H0}
\end{align}
where we have defined $\omega$ such that $2\omega$ is the energy
between ground state and biexciton, and $\omega+\delta$ is the energy
of the single exciton, with $\delta$ equal to half the biexciton
binding energy $\Delta E$. $\Omega(t)$ is the time-dependent coupling
cause by a laser pulse. We write the Hamiltonian in a frequency-modulated
rotating reference frame, with $\Omega(t)=\Omega_{t}\exp\left(i(\omega-\Delta)t-i\phi_{t}\right)$
(where $\Delta\equiv$detuning) , as
\begin{equation}
H=\left(\begin{array}{ccc}
-\Delta-\dot{\phi_{t}} & \Omega_{t} & 0\\
\Omega_{t} & \delta & \Omega_{t}\\
0 & \Omega_{t} & \Delta+\dot{\phi_{t}}\end{array}\right).\label{eq:Hmod}
\end{equation}
Here the basis is \{$e^{-i\left[(\omega-\Delta)t-\phi\right]}\left|g\right\rangle $,
$\left|X\right\rangle$,
$e^{i\left[(\omega-\Delta)t-\phi\right]}\left|XX\right\rangle$\}.

When $\Omega_{t}=0$, the eigenvectors of $H$ are the three basis
states, with eigenvalues \{$-\dot{\phi_{t}}$, $\delta$, $\dot{\phi_{t}}$\}.
We envisage that when $\dot{\phi_{t}}$ sweeps from negative to positive,
the eigenvector would change from $\left|g\right\rangle $ to $\left|XX\right\rangle $,
following which the system would be excited, by-passing the intermediate
state adiabatically, if a pulse, being sufficiently slow-varying,
is applied to induce the level anti-crossing.

In the following, we choose the following specific functional forms
for $\Omega_{t}$ and $\dot{\phi}_{t}$~\cite{goswami2003ops}
\begin{subequations}
\begin{eqnarray}
\Omega_{t} & = & A\textrm{sech}(\alpha t),\\
\dot{\phi_{t}} & = & \mu\alpha\tanh(\alpha t),\label{eq:pulseshape}
\end{eqnarray}
\end{subequations}
as shown in Fig.~\ref{fig:adiabatic}(a). Using
these waveforms, the adiabatic eigenvalues of the Hamiltonian are
computed and plotted in Fig.~\ref{fig:adiabatic}(b). As expected,
the eigenvectors $\left|g\right\rangle $ and $\left|XX\right\rangle $
exchange with each other. To illustrate the level anti-crossing, the
eigenvalues for the case of no interactions ($\Omega_{t}=0$) is also
plotted in dashed lines. We see that the single-exciton state does
not participate in the level crossing. Thus it can be inferred that
the occupation of $\left|X\right\rangle $ would be kept low because
the third eigenvalue $\delta$ is separated from the remaining two
eigenvalues; the larger $\delta$, the lower would be the occupation
of $\left|X\right\rangle $.

The square of components of eigenvectors for the middle eigenvalue
are plotted in solid lines in Fig.~\ref{fig:adiabatic}(c). As this
eigenvector changes from initially $\left|g\right\rangle $ to finally
$\left|XX\right\rangle $, we expect this to be followed by the actual
physical system, if the pulse is sufficiently slow-varying.

It should be pointed out that although we have chosen specific waveforms
in Eq.~(\ref{eq:pulseshape}) for the pulse shape, other choices are also
possible, provided that {}``anti-crossing'' similar to Fig.~\ref{fig:adiabatic}
can be produced. For instance, Gaussian shape for $\Omega_{t}$ and
linear frequency sweep could also be used~\cite{goswami2003ops}.
However, waveforms in Eq.~(\ref{eq:pulseshape}) shows better adiabaticity,
and is used in the simulation.

\begin{figure}
\begin{centering}
(a)\includegraphics[width=7cm]{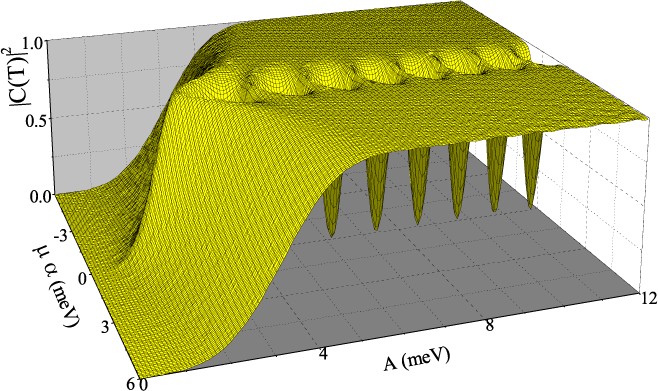}
\vskip 0.5cm
(b)\includegraphics[width=7cm]{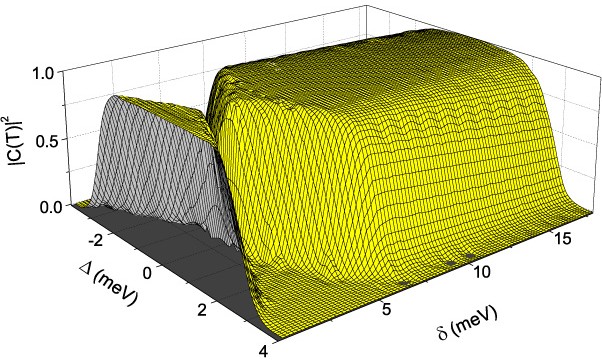}
\end{centering}
\caption{(Color online) Final population of the biexciton state. Here
$\alpha=0.6~{\textrm{meV}}\approx 1$~ps$^{-1}$
and $T=8~{\textrm{meV}}^{-1}\approx 5$~ps, which are determined
from Fig.~\ref{fig:adiabatic} with $\delta=10$~meV, consistent
with the binding energy of CdSe/ZnSe QDs.\quad{}(a) Fixing $\delta=10$~meV
and $\Delta=0$, $A$ and $\mu$ (i.e. $\mu\alpha$) are varied. If
$A$ is too small, the process fails to be adiabatic. Alternatively,
if $\mu\alpha$ is too small, no anti-crossing phenomena could be
observed.\quad{}(b) Fixing $A=6$~meV and $\mu\alpha=3$~meV,
detuning $\Delta$ and binding energy $2\delta$ are varied. The result
is acceptable for detuning within $\pm2$~meV.}
\label{fig:Performance}
\end{figure}

\section{Simulations\label{sec:EVOLUTION}}

The adiabaticity can be kept in two ways, by increasing the duration
of process or the pulse amplitude $A$. The occupation of intermediate
state can also be suppressed by increasing the duration. However,
long duration is an undesirable parameter in experiment because of
dephasing. In the following we fix a duration of $t\in\left[-T,T\right]$,
and investigate the adiabaticity of the state transfer as well as
the intermediate state population. We numerically solve the evolution
$\Psi(t)=a(t)\left|g\right\rangle +b(t)\left|X\right\rangle +c(t)\left|XX\right\rangle $,
with initial conditions $\Psi(0)=\left|g\right\rangle $. An example
is given by the solid lines Fig.~\ref{fig:adiabatic}(c). We see
that the actual evolution follows adiabatic approximation closely.
Note that with $\delta\approx10$~meV in the case of CdSe/ZnSe
QDs~\cite{PhysRevB.68.125316,kulakovskii1999fsb}, the duration $2T\approx10$~ps,
which is much shorter than the exciton dephasing time.

\begin{figure}[b]
\begin{centering}
\includegraphics[width=7cm]{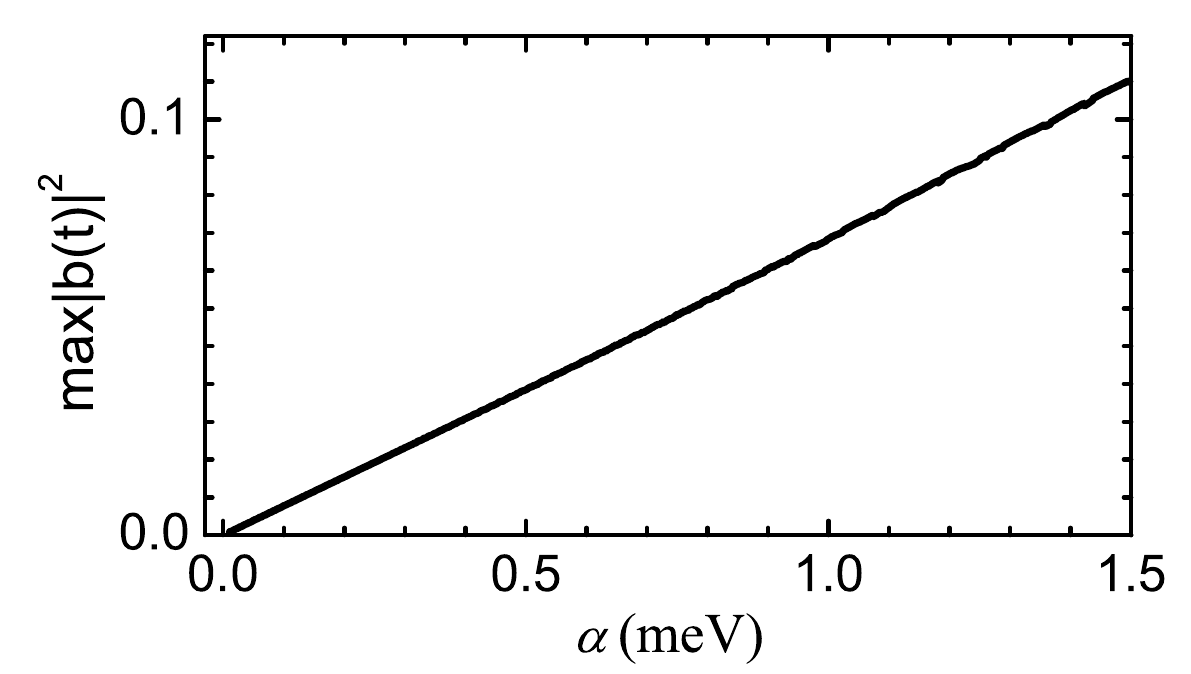}
\end{centering}
\caption{$\mu=2.4$ and $\delta=10$ meV are fixed, while $A$ is
adjusted accordingly such that for each $\alpha$ the final
biexciton population is at least $0.99$ and $\max|b(t)|^{2}$ is
minimized. Roughly a linear correlation is demonstrated, until
$\alpha$ is too large.
\label{fig:alpha}}
\end{figure}

To investigate on the dependence on the parameters, we plot the final
biexciton population $|c(T)|^{2}$ as a function of $A$ (coupling
magnitude) and $\mu\alpha$ (frequency-sweeping amplitude) in Fig.~\ref{fig:Performance}(a),
$\delta$ (binding energy/2) and $\Delta$ (detuning) in Fig.~\ref{fig:Performance}(b).
These plots have some foreseeable characteristics. The case $\mu\alpha\approx0$
corresponds to the usual case of two-photon Rabi oscillation, in which
the population transfer depends sensitively on the pulse area $A$.
This is demonstrated in the peaks and troughs along $\mu\alpha=0$
in Fig.~\ref{fig:Performance}(a), which are smoothed out as frequency-sweeping
is introduced. It is optimal for zero detuning, but some deviation
of $\Delta$ can be accepted. Uncertainty in $\delta$ (as large as
$\pm2$~meV) would not affect the efficacy of this process, either.

From the figures we remark that in contrast to the processes of Rabi
oscillation, this process is largely independent of the pulse area
($A$ and $\alpha$) and even the pulse shape. This is an experimentally
crucial feature, as the control over pulse area is often inexact under
realistic conditions, which would make the transferred population
lower than expected as in the ordinary two-photon Rabi flop scheme.

It is also of interest to investigate on the relation between time
duration $T\sim\alpha^{-1}$ in Eq.~(\ref{eq:pulseshape}). and the
single-exciton intermediate population $\max|b(t)|^{2}$. In Fig.~\ref{fig:alpha}
we plot $\max|b(t)|^{2}$ as a function of $\alpha$; where for each
$\alpha$ we use large enough $A$ such that the transferred population
($\left|g\right\rangle \rightarrow\left|XX\right\rangle $) is larger
than $0.99$ while $\max|b(t)|^{2}$ is minimized. We see a near-linear
correlation. Physically we understand that when we limit the process
to complete in shorter interval, the process becomes simply a transfer
via real excitation of the intermediate state
($\left|g\right\rangle \rightarrow\left|X\right\rangle \rightarrow\left|XX\right\rangle $).

\section{Effects of Dephasing}
\label{sec:effects-of-dephasing}

\begin{figure}[t]
\begin{centering}
\includegraphics[width=\columnwidth]{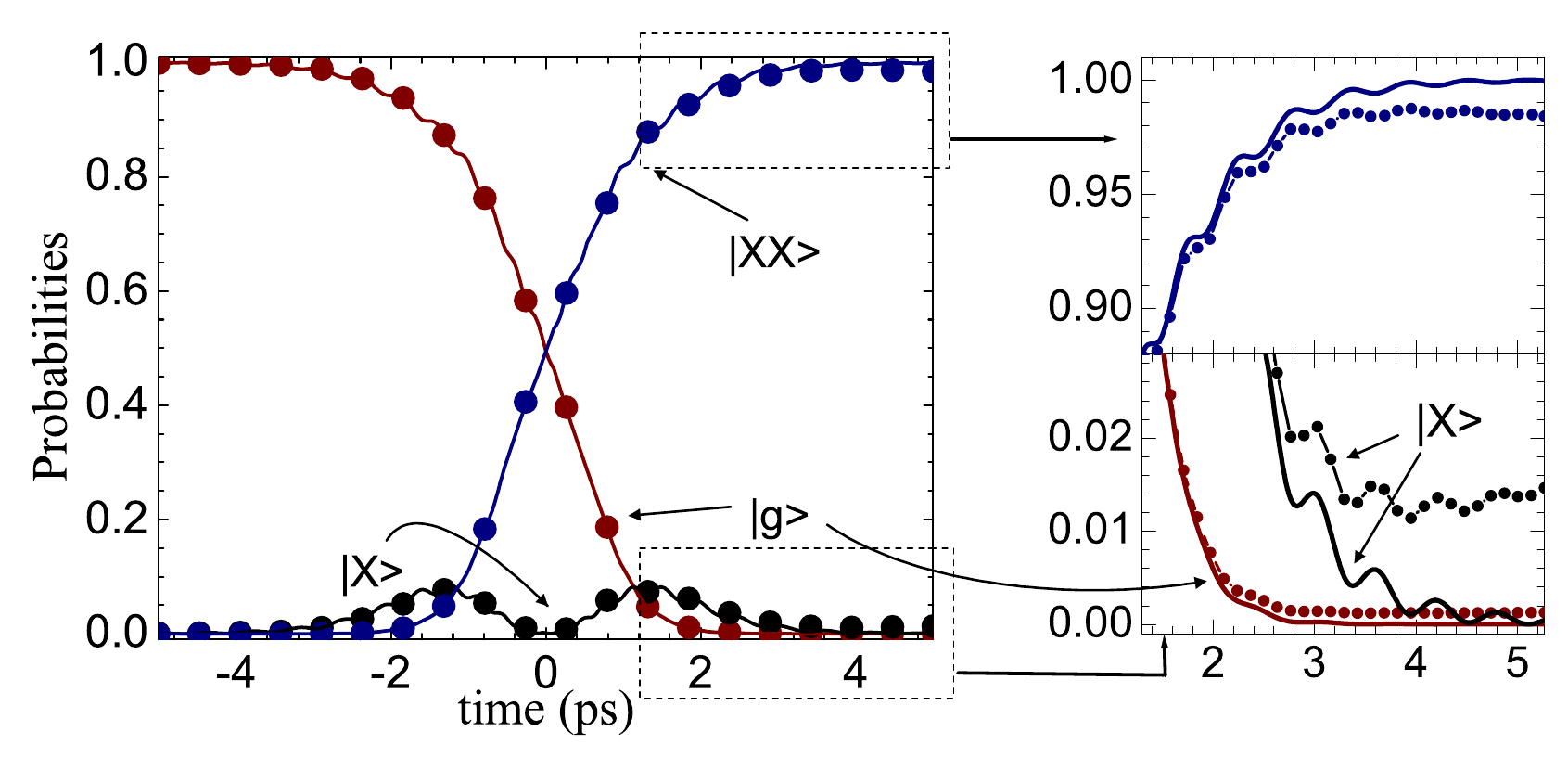}
(a) \vskip 0.3cm (b)\includegraphics[width=7cm]{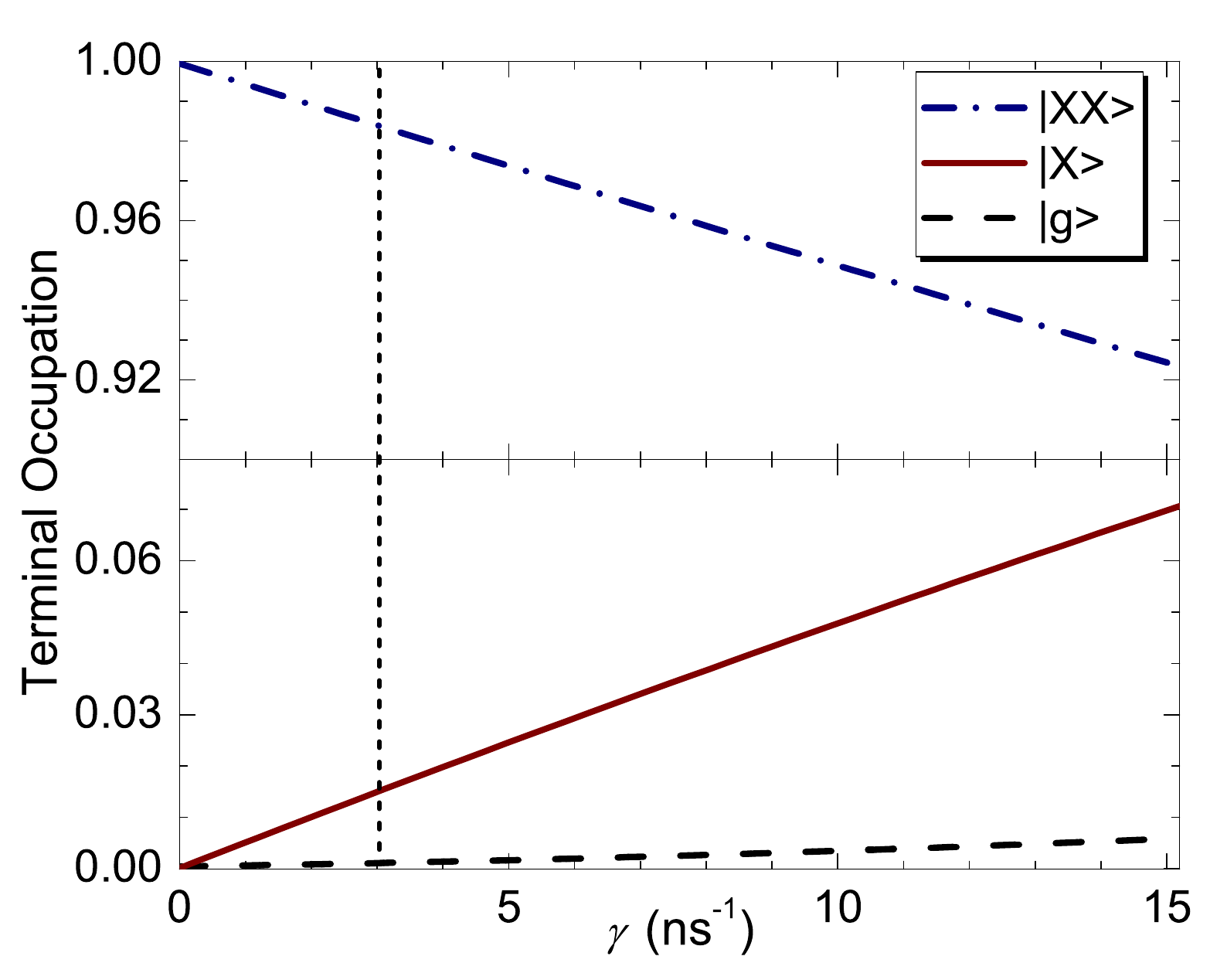}
\end{centering}
\caption{(Color online) (a) The numerical solution for the evolution of states
in CdSe/ZnSe QDs, for the case of no dephasing (solid lines) and the
case with dephasing (dots).\quad{}(b) The terminal state populations
for different values of $\gamma_{ij}$. The vertical dotted line indicates
the case of (a), $3.0$~ns$^{-1}$}
\label{fig:gamma}
\end{figure}

The analysis so far is less realistic in that we have neglected the
effect of dephasing present in QDs, which generally drives pure state
into mixed state. We thus consider the relaxation and dephasing of
excitons and biexcitons, which may be caused by spontaneous emission
and electron-phonon scattering~\cite{villasboas2005dro,Forstner2003}.
At low-temperatures, we consider just the spontaneous emission as
the limiting factor of the quantum operation~\cite{Palinginis2004}.
The spontaneous emission is modelled by an additional Lindblad term
in the master equation for density matrix $\rho$:
\begin{equation}
\partial_{t}\rho=-i[H_{0},\rho]+L(\rho),\label{eq:LindForm}
\end{equation}
where the Lindblad super-operator $L$ is defined by:\begin{equation}
L(\rho)=\sum_{ij}\frac{\gamma_{ij}}{2}\left[2\sigma_{ij}^{\dagger}\rho\sigma_{ij}-\sigma_{ij}\sigma_{ij}^{\dagger}\rho-\rho\sigma_{ij}\sigma_{ij}^{\dagger}\right],\label{eq:ExplicitLind}\end{equation}
with $\left\{ i,j\right\} =\left\{ \left|XX\right\rangle ,\left|X\right\rangle \right\} $,
or $\left\{ \left|X\right\rangle ,\left|g\right\rangle \right\} $,
signifying the transition from $i$ to $j$.

In the case of CdSe/ZnSe QDs, in the elimination of electron-phonon
interactions at low temperatures, it was determined to be $\gamma_{ij}\approx3.0$~ns$^{-1}$~\cite{Flissikowski2001}.
Together with $\delta=10$~meV and the pulse shape of Eq.~(\ref{eq:pulseshape}), the evolution of different
state populations are plotted in Fig.~\ref{fig:gamma}. It shows
only a slight reduction of the final population of $\left|XX\right\rangle $,
while those of $\left|X\right\rangle $ and $\left|g\right\rangle $
increase.

\section{Conclusion}

We have demonstrated the process of biexciton creation which is robust
against uncertainties in parameters of the system and the controlling
pulse, and keeps the involvement of single exciton state low. Using
a specific pulse shape, we showed the physical system adiabatically
follows the eigenvector. By plotting the final biexciton population
against various parameters, we have demonstrated the efficacy of the
proposed process does not have sensitive dependence on the pulse area,
pulse duration, level position, and detuning, which is the case for
the two-photon Rabi flop scheme. The maximum occupation of single
exciton state has an approximate linear relationship to the inverse
duration of the process, which means that the single-exciton state
could be kept low provided that sufficient time is given. Using the
experimental dephasing rate for CdSe/ZnSe quantum dots, we showed
that dephasing only causes a slight reduction in efficiency. \vfill{}

\end{document}